\documentclass[12pt]{article}
\usepackage{cite}
\usepackage{graphicx}

\begin{document}
\newcommand{\ome}{\omega_{\rm rot}}
\newcommand{\sumA}{\sum_{i=1}^A}
\newcommand{\boldr}{\mbox{\boldmath$r$}}
\newcommand{\boldj}{\mbox{\boldmath$j$}}
\newcommand{\boldrho}{\mbox{\boldmath$\rho$}}

\title{
Superdeformed bands in neutron-rich Sulfur isotopes 
suggested by cranked Skyrme-Hartree-Fock calculations
}
\author{T. Inakura$^a$, S. Mizutori$^b$, M. Yamagami$^a$ 
and K. Matsuyanagi$^a$\\
{\small\it $^a$ Department of Physics, Graduate School of Science,}\\
{\small\it Kyoto University, Kitashirakawa, Kyoto 606-8502, Japan}\\
{\small\it $^b$ Department of Human Science, Kansai Women's College,}\\
{\small\it Kashiwara City, Osaka 582-0026, Japan}\\
}
\date{\today}
\maketitle

\begin{abstract}

On the basis of the cranked Skyrme-Hartree-Fock
calculations in the three-dimensional coordinate-mesh representation,
we suggest that, in addition to the well-known candidate $^{32}$S, 
the neutron-rich nucleus $^{36}$S and the drip-line nuclei,
$^{48}$S and $^{50}$S,
are also good candidates for finding
superdeformed rotational bands in Sulfur isotopes. 
Calculated density distributions for the superdeformed states 
in $^{48}$S and $^{50}$S exhibit superdeformed neutron skins.  

\noindent
PACS: 21.60-n; 21.60.Jz; 27.30.+t \\

\noindent
Keywords: Cranked Skyrme-Hartree-Fock method; Superdeformation;
Neutron-rich nuclei; High-spin state; Sulfur isotopes

\end{abstract}

\vspace{2cm}
\newpage
\section{Introduction}

Recently, superdeformed (SD) rotational bands
have been discovered in $^{36}$Ar, $^{40}$Ca and $^{44}$Ti
\cite{sve00,sve01a,sve01b,ide01,chi03,lea00}.
One of the interesting new features of them is that they
are built on excited $0^+$ states and observed up to high spin,
in contrast to the SD bands in heavier mass regions
where low-spin portions of them are unknown in almost all cases
\cite{nol88,abe90,jan91,bak95,bak97}.
These excited $0^+$ states may be associated with multiparticle-multihole
excitations from the spherical closed shells, so that
we can hope to learn from such data detailed relationships between 
spherical shell model and SD configurations. 
For the mass $A=$30-50 region, 
although existence of a SD band in $^{32}$S with
the SD magic number $N=Z=16$ has been expected for a long time
\cite{rag78},
it has not yet been observed and remains as a great challenge
\cite{dob98, yam00,mol00,rod00,tan01}.

In this paper, as a continuation of the systematic theoretical search
\cite{yam00,ina02a}
for SD bands in the mass $A=$30-50 region 
by means of the cranked Skyrme-Hartree-Fock (SHF) method\cite{ben03},
we would like to suggest that, in addition to $^{32}$S, 
the neutron-rich nucleus $^{36}$S and the nuclei, $^{48}$S and $^{50}$S,
which are situated close to the neutron-drip line\cite{wer94,wer96},
are also good candidates for finding
SD rotational bands in Sulfur isotopes.
The appearance of the SD band in $^{36}$S is suggested 
in connection with the SD shell structure at $N=20$ 
characterizing the observed SD band in $^{40}$Ca.
The drip-line nuclei, $^{48}$S and $^{50}$S, are expected to constitute a new 
``SD doubly closed'' region associated with the
SD magic numbers, $Z=16$ for protons and $N \simeq 32$ for neutrons. 
An interesting theoretical subject for the SD bands in nuclei
near the neutron drip line is to understand deformation properties of 
neutron skins. The calculated density distributions indeed exhibit 
superdeformed neutron skins. 

The calculation has been carried out 
with the use of the three-dimensional (3D), 
Cartesian coordinate-mesh representation 
without imposing any symmetry restriction\cite{yam00,ina02a}.
In parallel, we also carry out the standard calculations
\cite{dav80,bon85,bon87,taj01} 
imposing reflection symmetries. 
By comparing the symmetry restricted and unrestricted calculations,
we can examine the stability of the SD solutions of the
SHF equations against reflection-asymmetric deformations.
In this way, we have found several cases where the SD minima 
obtained in the symmetry-restricted calculations are in fact unstable
with respect to the reflection-asymmetric deformations.
In general, the SD states are rather soft against reflection-asymmetric 
deformations, so that we need careful study about the 
stabilities of them against various kinds of deformation breaking
the reflection symmetries. 

After a brief account of the cranked SHF calculational procedure
in Section~2, we present and discuss results of the calculation
in Section~3, and give conclusions in Section~4. 
We shall present deformation energy curves for Sulfur isotopes 
from $^{32}$S to $^{50}$S, and focus our attention on 
properties of rotational bands built on the SD $0^+$ states, 
stabilities of the SD local minima 
against the reflection-asymmetric deformations,
and density distributions of the SD states.

A preliminary version of this work was reported in \cite{ina02b}.

\section{Cranked SHF calculation}

Since the cranked SHF method in the 3D coordinate-mesh representation 
is well known\protect\cite{dav80,bon85,bon87,taj01}, 
we here give only a minimum description about 
the computational procedure actually adopted.
For a recent comprehensive review 
on selfconsistent mean-field models for nuclear structure,
see Ref. \cite{ben03}. 
The cranked SHF equation for a system uniformly rotating about the
$x$-axis is given by 

\begin{equation}
\delta<H - \ome \hat{J}_x>=0,
\end{equation}

\noindent
where $H$, $\ome$ and $\hat{J}_x$ mean the Hamiltonian with the
Skyrme interaction, the rotational frequency and
the $x$-component of angular momentum, respectively,  
and the bracket denotes the expectation value with respect to a
Slater determinantal state.
We solve the cranked SHF equation
by means of the imaginary-time evolution technique\cite{dav80}
in the 3D Cartesian-mesh representation. 
The algorithm of numerical calculation is the standard one
\protect\cite{dav80,bon85,bon87,taj01},
except that we allow for both reflection- and
axial-symmetry breakings. 
In this case, it is important to accurately
fulfill the center-of-mass and principal-axis conditions. 
This is done by means of the constrained HF procedure\cite{flo73}. 
We solved these equations inside the
sphere with radius $R$=10~fm and mesh size $h$=1~fm,
starting with various initial configurations.
The accuracy for evaluating deformation energies 
with this mesh size was carefully checked 
by Tajima\cite{taj01} and was found to be quite satisfactory.
When we make a detailed analysis of density distributions,
however, we use a smaller mesh size of $h=\frac{1}{3}$ fm.
In addition to the symmetry-unrestricted cranked SHF calculation, 
we also carry out symmetry-restricted
calculations imposing reflection symmetries about 
the $(x,y)$-, $(y,z)$- and $(z,x)$-planes. 
Below we call these symmetry-unrestricted and -restricted cranked
SHF versions ``unrestricted'' and ``restricted'' ones, respectively. 

Solutions of the cranked SHF equation give local minima 
in the deformation energy surface. 
In order to explore the deformation energy surface around 
these minima and draw 
deformation energy curves as functions of deformation parameters, 
we carry out the constrained HF procedure 
with relevant constraining operators\cite{flo73}.  
For the Skyrme interaction, we adopt the widely used three versions;
SIII~\cite{bei75}, SkM$^*$\cite{bar82} and SLy4~\cite{cha98}.

\section{Results of calculation}
\subsection{\it Deformation energy curves}

Figure 1 shows deformation energy curves   
for Sulfur isotopes from $^{32}$S to $^{50}$S
obtained with the use of the SIII interaction.
Solid lines with and without filled circles represent 
the results obtained by the
unrestricted and restricted versions, respectively.
The result of calculation indicates that the SD minima 
(with the quadrupole deformation parameter $\beta_2\approx 0.6$) 
appear in the neutron-rich nucleus $^{36}$S and 
the drip-line nuclei, $^{48}$S and $^{50}$S,
in addition to the well-known case of $^{32}$S.
As seen in Figs.~2 and 3, similar results are obtained for
the SkM$^*$ and SLy4 interactions,
except that the SD states in $^{48}$S is unstable against the
reflection-asymmetric deformation for the SLy4 interaction
(see subsection 3.3).

As discussed in Refs.\cite{yam00,mol00,rod00,tan01,ina02a}, 
the SD local minimum in $^{32}$S  corresponds to the doubly closed shell
configuration with respect to the SD magic number $Z=N=16$ 
and involves two protons and two neutrons 
in the down-sloping single-particle levels 
originating from the $f_{7/2}$ shell.
The SD local minimum in $^{36}$S results from
the coherent combination of the SD magic number, $Z=16$, and 
the neutron shell effects occurring at large deformation for
$N=20$. The latter shell effect has been confirmed recently by the
discovery of the SD rotation band in $^{40}$Ca
\cite{ide01,chi03}. 
The SD shell gap at $N=20$ is associated with the 4p-4h excitation 
from below the $N=20$ spherical closed shell to the $f_{7/2}$ shell.
Focusing our attention on the occupation numbers of such high-$j$ shells
and distinguishing protons($\pi$) and neutrons($\nu$),
these SD configurations in $^{32}$S and $^{36}$S
are denoted in Figs.~1-3  
as $f_{\pi}^2 f_{\nu}^2$ and $f_{\pi}^2 f_{\nu}^4$, respectively.

The SD local minima in the drip-line nuclei, $^{48}$S and $^{50}$S,
result from the coherent combination of the proton SD shell effect and 
the neutron shell effects occurring at superdeformation for $N=$32-34.
The neutron configurations in these SD states are similar
to those in the known SD bands in $^{60}$Zn and $^{62}$Zn
associated with the SD magic numbers $N=$30-32\cite{sve99,sve97}.
We find that the SD shell effect is strong also for $N=34$
in the Sulfur isotopes under consideration, while the SD local
minimum in $^{46}$S with $N=30$ is unstable
against the reflection-asymmetric deformation (see subsection 3.3).
In the drip-line nuclei $^{48}$S and $^{50}$S,
the $f_{7/2}$ shell is fully occupied even in the spherical limit and 
the SD configurations involve neutron excitations from
the $fp$-shell to the $g_{9/2}$ shell.
As before, focusing our attention on the occupation numbers 
of the high-$j$ shells,
let us use the notation $f_{\pi}^{n_1} g_{\nu}^{n_2}$ 
for a configuration in which single-particle levels originating from
the $f_{7/2}$ and $g_{9/2}$ shell
are occupied by $n_1$ protons and $n_2$ neutrons, respectively.
With such notations, the SD local minima in $^{48}$S and $^{50}$S
correspond to the $f_{\pi}^2 g_{\nu}^2$, and $f_{\pi}^2 g_{\nu}^4$
configurations, respectively.

The appearance of the SD minimum in $^{36}$S
suggests that we can expect a SD band associated with
the same neutron configuration to appear 
also in the $N=20$ isotone, $^{38}$Ar, 
situated between $^{36}$S and $^{40}$Ca.  
We examined this point and the result is shown in Fig.~4.
We find that the two local minima associated with the configurations
$f_{\pi}^2 f_{\nu}^2$ and $f_{\pi}^2 f_{\nu}^4$ 
compete in energy and their relative energy differs for
different versions of the Skyrme interaction:
As clearly seen in the deformation-energy curves obtained by the
symmetry-restricted calculations,
the former with smaller $\beta_2$ is slightly lower for SkM$^*$ and SLy4
while the latter with larger $\beta_2$ is slightly lower for SIII.
Counting both protons and neutrons, these local minima
respectively correspond to the 4p-6h and 6p-8h
configurations with respect to the spherical doubly closed shell 
of $^{40}$Ca. 
As we discussed in the previous papers\cite{ina02a,ina02b}, 
the two configurations can mix each other in the crossing region
through the reflection-symmetry breaking components in the mean field.
Specifically, around the crossing point between
the down-sloping $[321 \frac 32]$ level
(coming from the $f_{7/2}$ shell) and 
the up-sloping $[200 \frac 12]$ level
(coming from the $d_{3/2}$ shell below the $N=20$ spherical magic number),
the $r^3 Y_{31}$-type non-axial octupole deformation is generated,
and they mix each other through this component of the mean field 
(see Fig.~5).
Note that the matrix element of the operator $r^3 Y_{31}$
between the two levels satisfies the selection rules,  
$\Delta n_3=2$ and $\Delta \Lambda=1$,
for the asymptotic quantum numbers $n_3$ and $\Lambda$. 
As a result of this mixing,
the deformation-energy curve becomes rather flat
in the symmetry-unrestricted calculation.
Recently, the SD band corresponding to the 4p-6h configuration
was found in experiment\cite{rud02}.
The data suggest significant competition between 
different configurations, which requires 
further analysis of shape fluctuation dynamics 
by going beyond the static mean-field approximation.

\subsection{\it SD rotational bands}

Let us focus our attention on the SD local minima shown in Figs.~1-3, 
and investigate properties of the rotational bands built on them. 
Excitation energies of these SD rotational bands are plotted in Fig.~6 
as functions of angular momentum.
These rotational bands are calculated by cranking each SHF solution 
(the SD local minima in Figs.~1-3) 
and following the same configuration with increasing value of $\ome$
until the point where we cannot clearly identify the continuation
of the same configuration any more. 
Thus, the highest values of angular momentum in this figure 
does not necessarily indicate the band-termination points but 
merely suggest that drastic changes in their microscopic structure 
take place around there.
Different slopes with respect to the angular momentum
between $^{36}$S and $^{50}$S can be easily understood 
in terms of the well known scaling factor $A^{5/3}$ 
for the rigid-body moment of inertia.
This point can be confirmed in Fig.~7 which displays
the angular momentum $I$, 
the kinematical and dynamical moments of inertia, 
${\cal J}^{(1)}=I/\ome$ and ${\cal J}^{(2)}=dI/d\ome$,
and the rigid-body moments of inertia
${\cal J}_{\rm rig}=m\int \rho({\boldr})(y^2+z^2)d{\boldr}$
as functions of the rotational frequency $\ome$. 
We see that the calculated moments of inertia 
are slightly larger than the rigid-body values at $\ome=0$, 
and smoothly decrease as $\ome$ increases
until $\ome \approx$2.5 and 1.8 MeV$/\hbar$ for 
$^{32,36}$S and $^{50}$S, respectively.
The result calculated with the SLy4 interaction is shown here, 
but we obtained similar results also with the SIII and SkM$^*$ interactions.

Calculated quadrupole deformation parameters $(\beta_2,\gamma)$
are displayed in the upper portion of Fig.~6.
We see that the $\beta_2$ values slowly decrease while the axial-asymmetry
parameters $\gamma$ gradually increase with increasing angular momentum
for all cases of $^{32}$S, $^{36}$S and $^{50}$S.
The variations are rather mild
in the range of angular momentum shown in this figure.
Single-particle energy diagrams (Routhians) for these SD bands 
are displayed in Fig.~8 as functions of the rotational frequency $\ome$. 
This figure indicates that level crossings take place
in $^{36}$S and $^{50}$S if we further increase the angular momentum.

\subsection{\it Stabilities of the SD states against 
reflection-asymmetric deformations}

Let us examine stabilities of the SD local minimum
against both the axially symmetric and asymmetric octupole deformation
($Y_{30},Y_{31},Y_{32},Y_{33}$).
Figure 9 presents deformation energy curves as functions of 
the octupole deformation parameters  $\beta_{3m}(m=0,1,2,3)$  
for fixed quadrupole deformation parameters
(the equilibrium  value of $\beta_2$ at the SD minimum in each nucleus 
and $\gamma=0$).
The computation was carried out by means of the constrained HF procedure 
with the use of the SIII, SkM$^*$, and SLy4 interactions.
The result of calculation clearly indicates that 
the SD states in $^{32}$S, $^{36}$S and $^{50}$S 
are stable against the octupole deformations 
and that they are softer for $\beta_{3m}$ 
with lower values of $m$ (i.e., for $\beta_{30}$ and $\beta_{31}$),
irrespective of the Skyrme interactions used.
We obtaind a similar result also for $^{48}$S (but omitted in this figure).

Although the SD minima in $^{32}$S, $^{36}$S and $^{50}$S are stable 
with respect to $\beta_{3m}(m=0,1,2,3)$, we found several cases where the
SD minima obtained in the symmetry-restricted calculations become
unstable when we allow for reflection-asymmetric deformations 
of a more general type. 
As a first example, let us discuss the SD minimum in $^{46}$S
which appears in the restricted calculation (see Figs.~1-3).
In this case, the coupling between the down-sloping  
$[330 \frac{1}{2}]$ level (associated with the $f_{7/2}$ shell)
and the up-sloping $[202 \frac{5}{2}]$ level 
(stemming from the $d_{5/2}$ shell)  
takes place in the proton configuration,
when we allow for the breaking of both the axial and reflection symmetries.
Thus, the SD configuration $f_{\pi}^2 g_{\nu}^2$  
mixes with the $g_{\nu}^2$ configuration (which lacks the proton excitation
to the $f_{7/2}$ shell and has a smaller equilibrium value of $\beta_2$). 
As a consequence of this mixing, the barrier between the two
configurations disappears and the SD minimum becomes unstable
in the unrestricted calculations (see Figs.~1-3).  
Note that the difference $\Delta n_3$ 
in the asymptotic quantum number $n_3$ between the 
two single-particle levels, $[330 \frac{1}{2}]$ and $[202 \frac{5}{2}]$,
is three, so that they cannot be mixed by the octupole operator $r^3Y_{32}$ 
which transfers the asymptotic quantum numbers $n_3$ and $\Lambda$ by
$\Delta n_3=1$ and $\Delta \Lambda=2$.
Thus, this mixing may be associated with the reflection-asymmetric 
deformation of a more higher order like $r^5 Y_{52}$.

As a second example, we take up the SD minimum in $^{48}$S.
In this case, two configurations,
$f_{\pi}^2 g_{\nu}^2$ and $f_{\pi}^2 g_{\nu}^4$ 
compete in energy and their relative energy differs for
different versions of the Skyrme interaction (see Figs.~1-3).
When we allow for the breaking of both the axial and reflection symmetries,
the coupling between the down-sloping  
$[431 \frac{3}{2}]$ level (associated with the $g_{9/2}$ shell)
and the $[310 \frac{1}{2}]$ level in the $fp$ shell  
takes place in the neutron configuration,
so that they mix each other.
Note that the $[431 \frac{3}{2}]$  and $[310 \frac{1}{2}]$ levels
satisfy the selection rules, $\Delta n_3=2$ and $\Delta \Lambda=1$,
for the matrix elements of the octupole operator $r^3 Y_{31}$.
In the calculation with the SLy4 interaction, 
since the former configuration with smaller $\beta_2$ 
is situated slightly lower in energy than the latter, 
the barrier between the two configurations disappears 
as a result of this mixing.
This mixing effect in conjunction with that mentioned above 
for the $f_{\pi}^2 g_{\nu}^2$ configuration in $^{46}$S
deteriorates the SD minimum for the SLy4 case.

The above examples indicate detailed microscopic mechanisms
within the mean-field theory
how the stability of the SD local minimum is determined 
by relative energies between the neighboring configurations
and their mixing properties.

\subsection{\it Density distributions}

Figure 10 displays the neutron and proton density profiles for
the SD states in $^{32}$S, $^{36}$S and $^{50}$S
calculated with the use of the SLy4 interaction.
We obtained similar results also for SIII and SkM$^*$.
In this figure, equi-density lines with 50\% and 1\% of
the central density in the $(x,y)$- and $(y,z)$-planes are drawn 
for the SD bands at $I=0$ and at high spins.
We can clearly see that superdeformed neutron skin appears in $^{50}$S 
which is situated close to the neutron drip line.
The root-mean-square values,
$\sqrt{\langle x^2 \rangle}, \sqrt{\langle y^2 \rangle},
\sqrt{\langle z^2 \rangle}$ 
and $R_{\rm rms}=\sqrt{\langle \boldr^2 \rangle}$,
of these density distributions are listed in Table 1.  
To indicate the deformation properties 
of the neutron skin in $^{50}$S, calculated values for protons and neutrons
are separately listed together with their sums and differences.
We obtained density distributions similar to those for $^{50}$S 
also for the SD state in $^{48}$S.
A similar result of theoretical calculation 
exhibiting the superdeformed neutron skin
was previously reported in Ref.\cite{ham95}
for the SD state in the very neutron-rich nucleus $_{~66}^{208}$Dy$_{142}$.  

\section{Conclusions}

On the basis of the cranked SHF
calculations in the 3D coordinate-mesh representation,
we have suggested that, in addition to the well-known candidate $^{32}$S, 
the neutron-rich $^{36}$S and 
the the drip-line nuclei, $^{48}$S and $^{50}$S, 
are also good candidates for finding
SD rotational bands in Sulfur isotopes. 
Calculated density distributions for the SD states 
in $^{48}$S and $^{50}$S, which are situated close to the neutron-drip line, 
exhibit superdeformed neutron skins. 

\section*{Acknowledgements}
The numerical calculations were performed on the NEC SX-5 supercomputers
at RCNP, Osaka University, 
and at Yukawa Institute for Theoretical Physics, Kyoto University.
This work was supported by the Grant-in-Aid  for Scientific
Research (No. 13640281) from the Japan Society for the Promotion of Science.

%\newpage

%\end{document

\newpage
\vspace{2cm}
\begin{center}
Table 1\\
\end{center}

Root-mean-square values,
$\sqrt{\langle x^2 \rangle}, \sqrt{\langle y^2 \rangle},
\sqrt{\langle z^2 \rangle}$ 
and $R_{\rm rms}=\sqrt{\langle \boldr^2 \rangle}$, 
of the density distributions at $I=0$ (second column) 
and at $I=20, 22, 28$ (third column)
of the SD band in $^{32}$S, $^{36}$S and $^{50}$S,
calculated with the use of the SLy4 interaction.
Neutron and proton contributions are separately listed
together with their sums (total) and differences (diff.).\\

\begin{center}
\begin{tabular}{|c|cccc|cccc|}
\hline
\raisebox{-1.25em}[0pt][0pt]{{\LARGE $^{32}$S}} & \multicolumn{4}{c|}{$I=0$} &  \multicolumn{4}{c|}{$I \sim 20$}\\ \cline{2-9}
&  $\sqrt{ \langle x^2 \rangle }$ & $\sqrt{ \langle y^2 \rangle }$ & $\sqrt{ \langle z^2 \rangle }$ & $R_{\mathrm{rms}}$
&  $\sqrt{ \langle x^2 \rangle }$ & $\sqrt{ \langle y^2 \rangle }$ & $\sqrt{ \langle z^2 \rangle }$ & $R_{\mathrm{rms}}$\\ \hline
total      &  1.53 &   1.53 &  2.85  &  3.57  &  1.53 &  1.67 &  2.67  &  3.50 \\
neutrons   &  1.52 &   1.52 &  2.83  &  3.55  &  1.52 &  1.66 &  2.65  &  3.48 \\
protons    &  1.54 &   1.54 &  2.86  &  3.60  &  1.54 &  1.68 &  2.68  &  3.52 \\ \hline
diff.      & -0.02 &  -0.02 & -0.04  & -0.04  & -0.02 & -0.02 & -0.04  & -0.05 \\
\hline
\hline
\raisebox{-1.25em}[0pt][0pt]{{\LARGE $^{36}$S}} & \multicolumn{4}{c|}{$I=0$} &  \multicolumn{4}{c|}{$I \sim 22$}\\ \cline{2-9}
&  $\sqrt{ \langle x^2 \rangle }$ & $\sqrt{ \langle y^2 \rangle }$ & $\sqrt{ \langle z^2 \rangle }$ & $R_{\mathrm{rms}}$
&  $\sqrt{ \langle x^2 \rangle }$ & $\sqrt{ \langle y^2 \rangle }$ & $\sqrt{ \langle z^2 \rangle }$ & $R_{\mathrm{rms}}$\\ \hline
total     & 1.59 &  1.59 &  2.78   &  3.58 & 1.61 &   1.73 &  2.63  & 3.53 \\
neutrons  & 1.62 &  1.62 &  2.78   &  3.60 & 1.64 &   1.75 &  2.64  & 3.57 \\
protons   & 1.55 &  1.55 &  2.78   &  3.54 & 1.57 &   1.69 &  2.61  & 3.48 \\ \hline
diff.     & 0.07 &  0.07 &  0.00   &  0.06 & 0.08 &   0.06 &  0.03  & 0.09 \\
\hline
\hline
\raisebox{-1.25em}[0pt][0pt]{{\LARGE $^{50}$S}} & \multicolumn{4}{c|}{$I=0$} &  \multicolumn{4}{c|}{$I \sim 28$}\\ \cline{2-9}
 &  $\sqrt{ \langle x^2 \rangle }$ & $\sqrt{ \langle y^2 \rangle }$ & $\sqrt{ \langle z^2 \rangle }$ & $R_{\mathrm{rms}}$
&  $\sqrt{ \langle x^2 \rangle }$ & $\sqrt{ \langle y^2 \rangle }$ & $\sqrt{ \langle z^2 \rangle }$ & $R_{\mathrm{rms}}$\\ \hline
total     & 1.81 &  1.81 &  3.11  &  4.03 & 1.82 &   1.96 &  2.95  &  3.98 \\
neutrons  & 1.90 &  1.90 &  3.17  &  4.16 & 1.91 &   2.05 &  3.02  &  4.12 \\
protons   & 1.62 &  1.62 &  2.96  &  3.75 & 1.63 &   1.75 &  2.79  &  3.67 \\ \hline
diff.     & 0.28 &  0.28 &  0.20  &  0.41 & 0.28 &   0.31 &  0.23  &  0.45 \\
\hline
\end{tabular}
\end{center}

\newpage

\begin{center}
\begin{figure}
\includegraphics[width=1.00\textwidth,keepaspectratio]{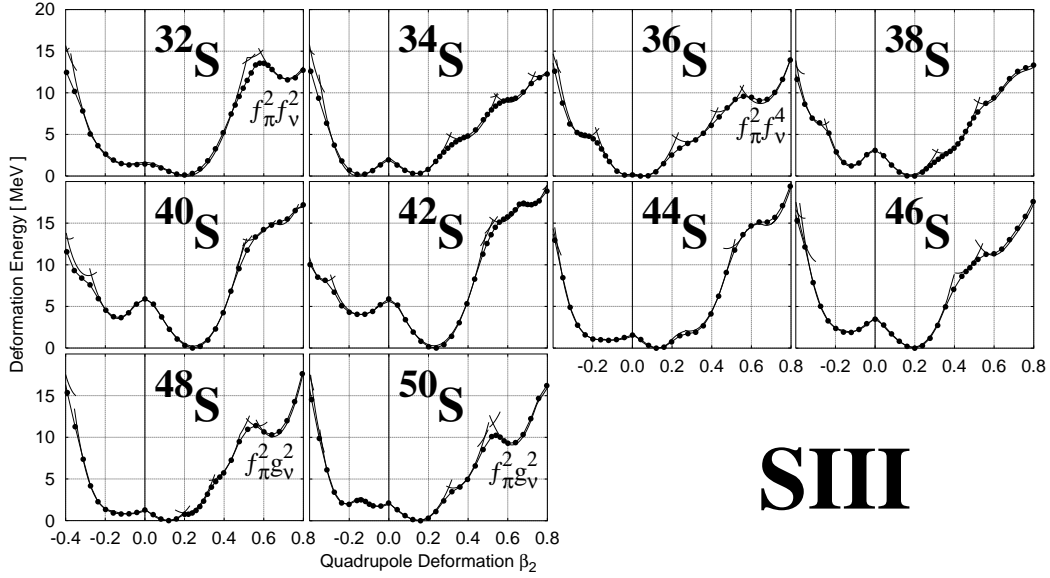}
\caption{\small
Deformation energy curves for Sulfur isotopes 
from $^{32}$S to $^{50}$S calculated at $I=0$ 
as functions of the quadrupole deformation $\beta_2$ 
by means of the constrained SHF procedure with
the SIII interaction. 
The deformation parameter is defined as
$\beta_2=\frac {4\pi}{5}~\langle\sumA r_i^2 Y_{20}(\theta_i,\phi_i)\rangle 
/\langle\sumA \boldr_i^2\rangle.$
The axial-asymmetry parameter $\gamma$ is constrained to be zero.
Solid curves with and without filled circles represent 
the results obtained by the
unrestricted and restricted versions, respectively.
The notation $f_{\pi}^{n_1} f_{\nu}^{n_2}$  
indicates a configuration in which single-particle levels 
originating from the $f_{7/2}$ shell are occupied by $n_1$ protons 
and $n_2$ neutrons.
Likewise, $f_{\pi}^{n_1} g_{\nu}^{n_2}$
indicates that levels from the $f_{7/2}$ shell are occupied by $n_1$ protons 
and those from the $g_{9/2}$ shell by $n_2$ neutrons.
}
\end{figure}
\end{center}

\begin{center}
\begin{figure}
\includegraphics[width=1.00\textwidth,keepaspectratio]{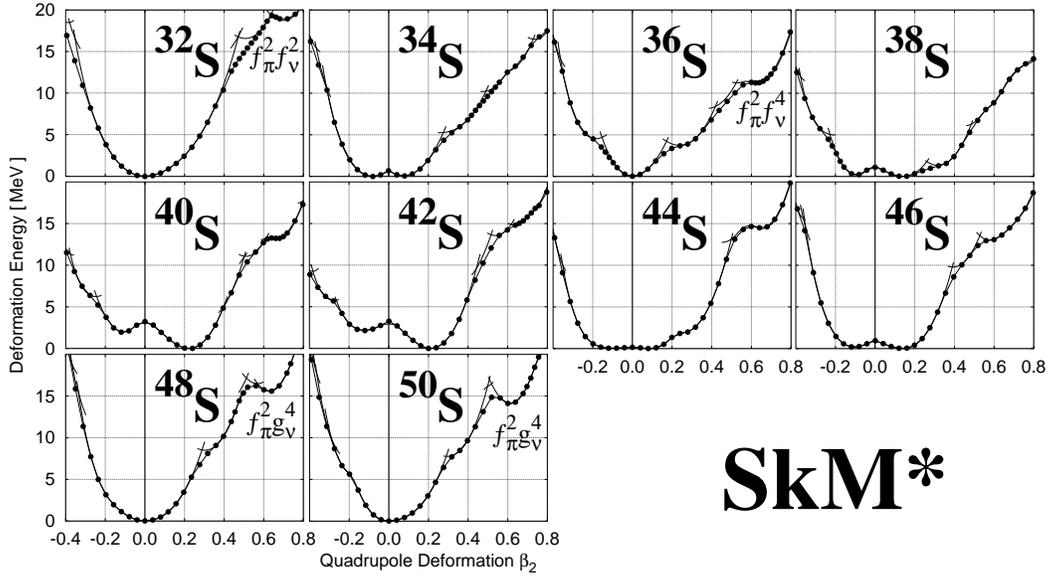}
\caption{\small 
The same as Fig.1 but for the SkM$^*$ interaction.
}
\end{figure}
\end{center}

\begin{center}
\begin{figure}
\includegraphics[width=1.00\textwidth,keepaspectratio]{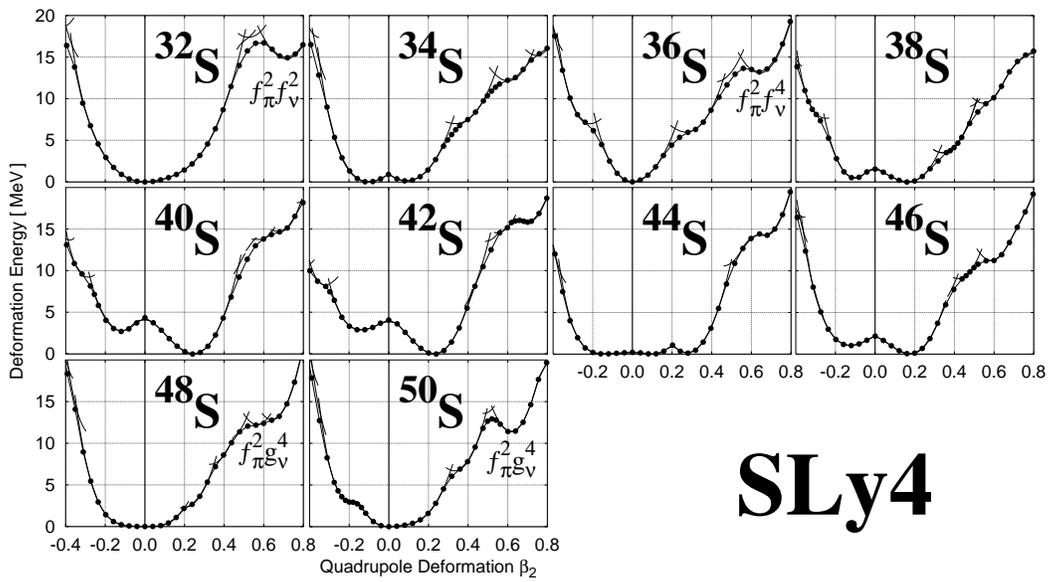}
\caption{\small
The same as Fig.1 but for the SLy4 interaction.
} 
\end{figure}
\end{center}

\begin{center}
\begin{figure}
\includegraphics[width=1.00\textwidth,keepaspectratio]{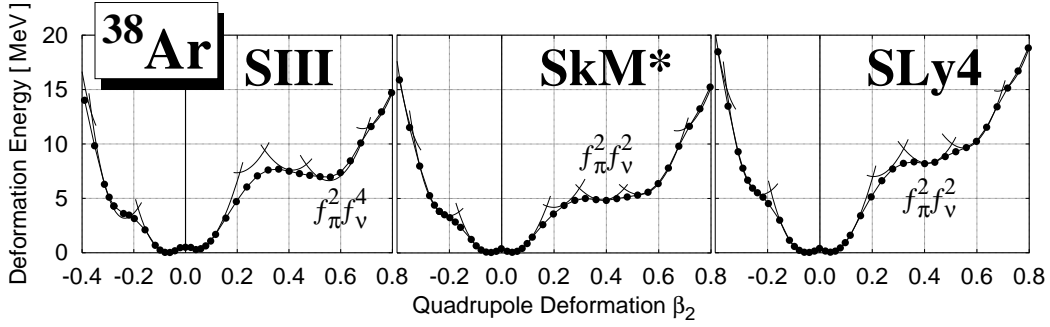}
\caption{\small 
The same as Fig.1 but for $^{38}$Ar
and for the SIII, SkM$^*$ and SLy4 interactions.
}
\end{figure}
\end{center}

\begin{center}
\begin{figure}
\includegraphics[width=1.00\textwidth,keepaspectratio]{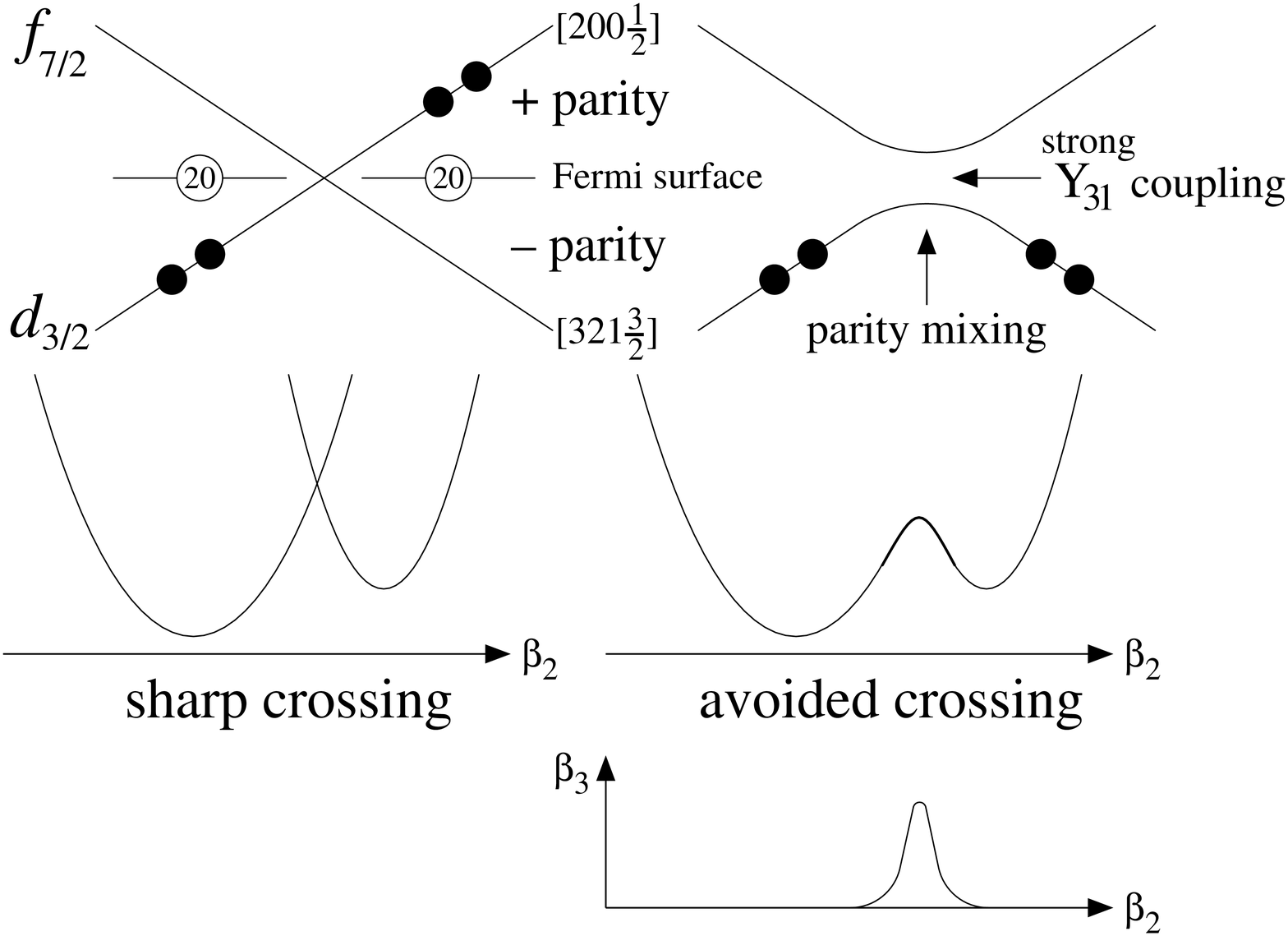}
\caption{\small 
Schematic illustration of configuration-mixing mechanism 
through the octupole components of the mean field.
When the reflection symmetry is imposed, 
the positive- and negative-parity single-particle levels sharply cross, 
and the two configurations 
(having different number of particles in the $f_{7/2}$ shell) 
do not mix within the mean-field approximation
(left-hand side).
In contrast, when such symmetry restriction is removed,
smooth crossover between the two configurations is possible 
via mixing of the positive- and negative-parity levels 
(right-hand side).
Octupole deformation $\beta_3$ of the mean field rises 
in the crossing region.
In this figure, the crossing between the two levels with the 
asymptotic quantum numbers $[321 \frac 32]$ and 
$[200 \frac 12]$ is illustrated as an example.
The two levels satisfy the selection rule, 
$\Delta n_3=2$ and $\Delta \Lambda=1$,
for the matrix elements of the non-axial octupole operator $r^3 Y_{31}$, 
so that the mixing between them takes place mainly 
through the $r^3 Y_{31}$ component of the mean field.
}
\end{figure}
\end{center}

\begin{center}
\begin{figure}
\includegraphics[width=1.00\textwidth,keepaspectratio]{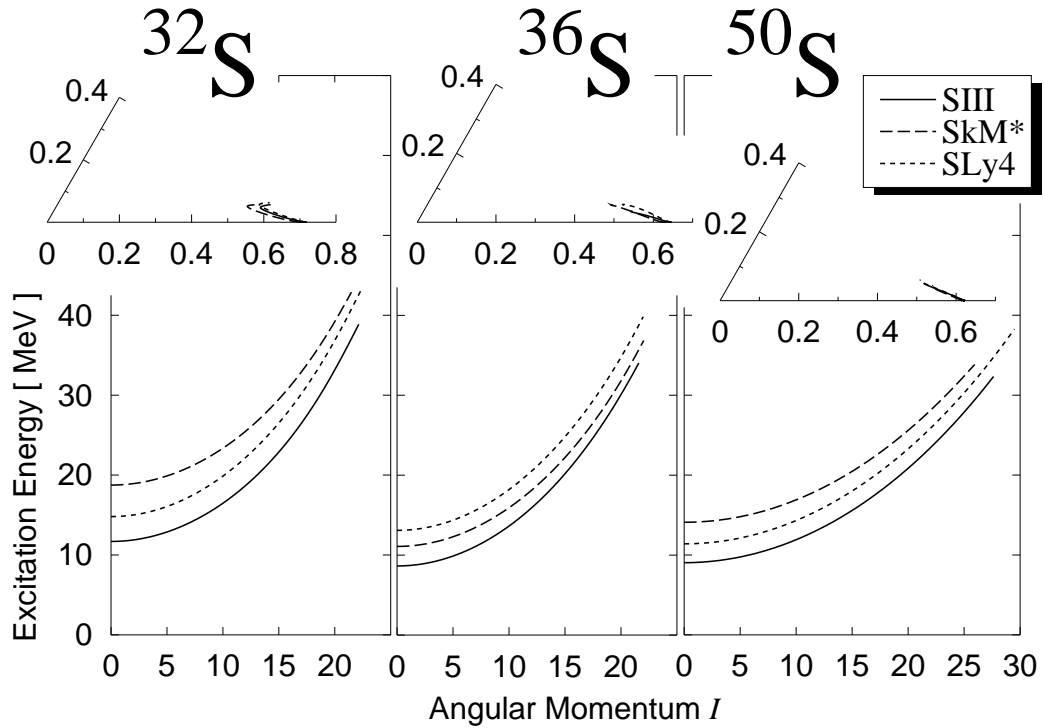}
\caption{\small
Plot of the excitation energies versus angular-momentum 
for the SD rotational bands in $^{32}$S, $^{36}$S, and $^{50}$S
calculated by means of the cranked SHF method.
Results obtained with the use of the SIII, SkM$^*$, and SLy4 interactions
are plotted by solid, dashed, and dotted curves, respectively.
Their shape evolutions as functions of angular momentum $I$
in the $(\beta_2,\gamma)$ plane are displayed in the upper portions.
The $\beta_2$ values decrease with increasing $I$.
}
\end{figure}
\end{center}

\begin{center}
\begin{figure}
\includegraphics[width=1.00\textwidth,keepaspectratio]{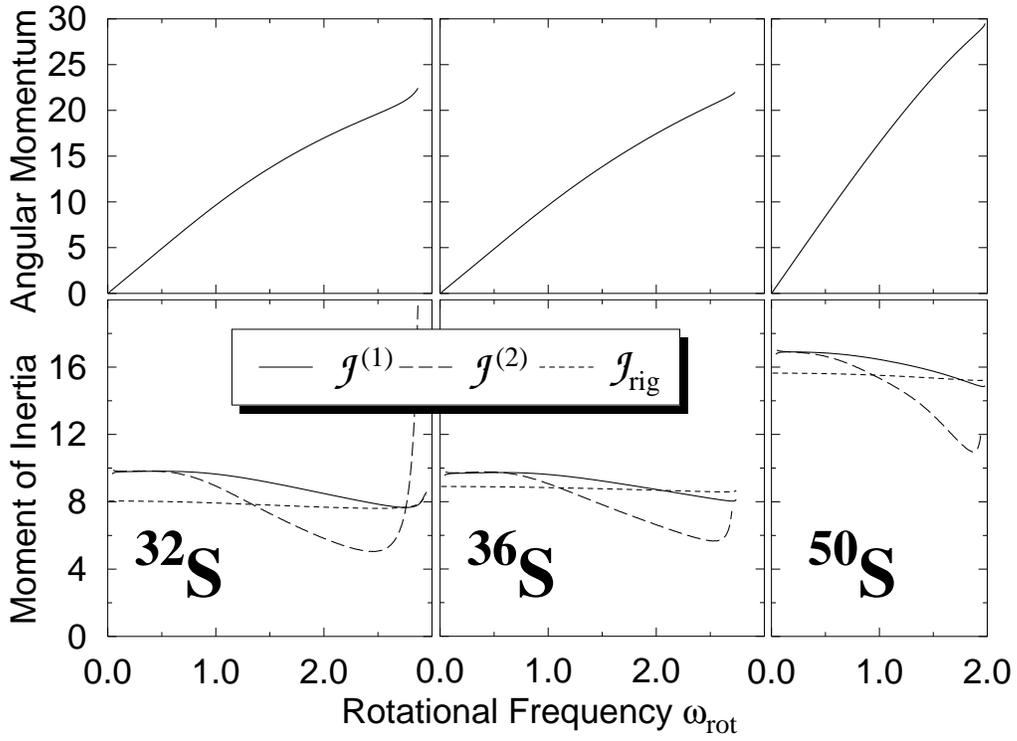}
\caption{\small
The angular momenta $I$ and the moments of inertia ${\cal J}$ 
are plotted as functions of rotational frequency $\ome$  
for the SD rotational bands in $^{32}$S, $^{36}$S, and $^{50}$S.
The SLy4 interaction is used.
Values of the kinematical and dynamical moments of inertia, 
${\cal J}^{(1)}=I/\ome$ and ${\cal J}^{(2)}=dI/d\ome$,
are plotted in unit of $\hbar^2$/MeV by solid and dashed curves, 
respectively.
For reference, the rigid-body moments of inertia
${\cal J}_{\rm rig}=m\int \rho({\boldr})(y^2+z^2)d{\boldr}$
evaluated with the calculated density $\rho({\boldr})$ 
are also indicated.
}
\end{figure}
\end{center}

\begin{center}
\begin{figure}
\includegraphics[width=1.00\textwidth,keepaspectratio]{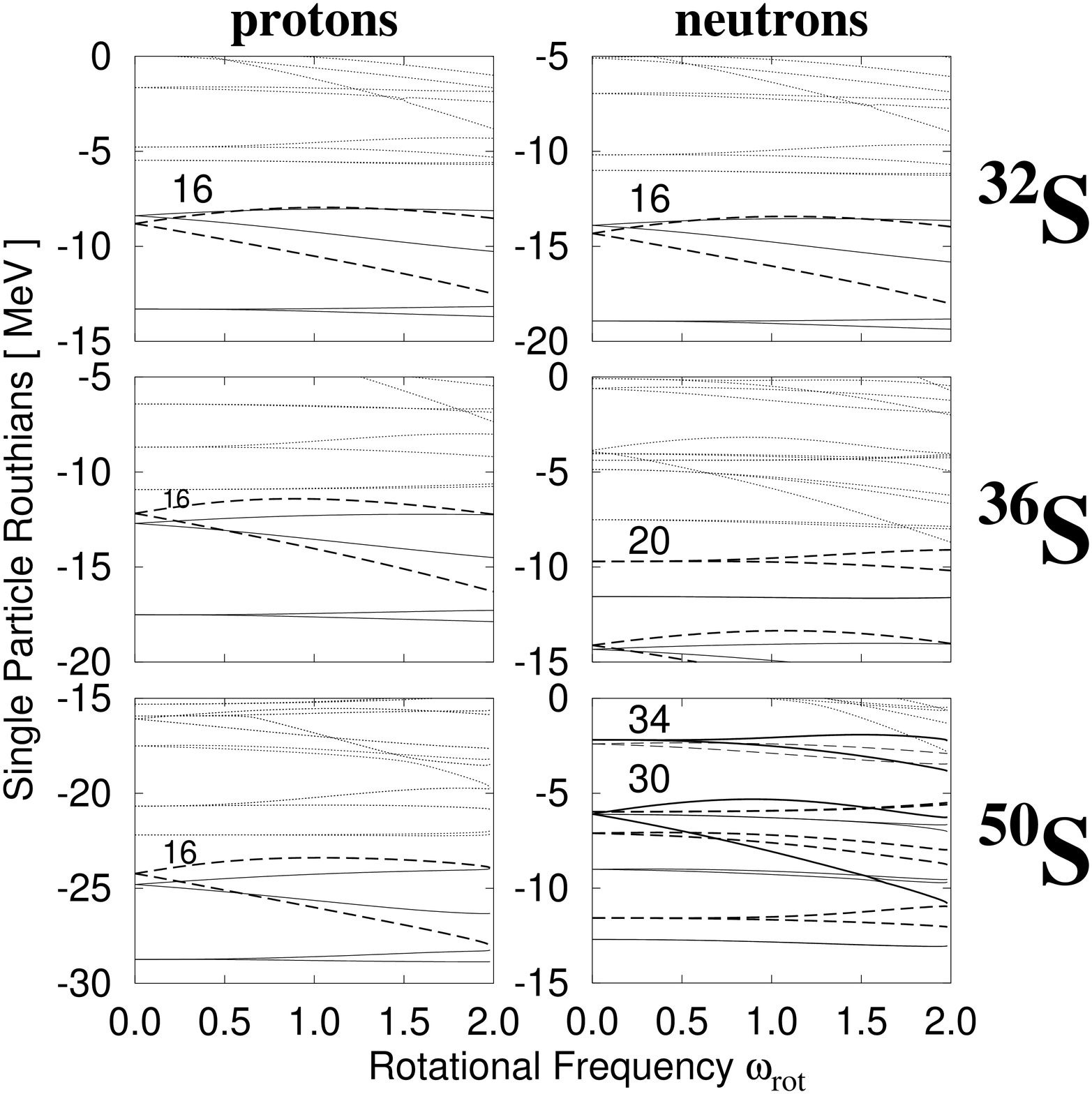}
\caption{\small
Single-particle energy diagrams (Routhians) for the SD bands 
in $^{32}$S, $^{36}$S, and $^{50}$S, 
plotted as functions of rotational frequency $\ome$.
The left(right)-hand side displays those for protons(neutrons).
The levels associated with the $g_{9/2}$ and $f_{7/2}$ shells
are drawn by thick-solid and thick-dashed lines, respectively.
Other occupied levels associated with the $sd$ and $fp$ shells 
are drawn by thin-solid and thin-dashed lines, respectively.
Unoccupied levels are drawn by thin-dotted lines.
Numbers indicate the Fermi surfaces and total numbers of 
single-particle states below them.
The result calculated with SLy4 is shown here, 
but we obtained similar results also with SIII and SkM$^*$.
}
\end{figure}
\end{center}

\begin{center}
\begin{figure}
\includegraphics[width=1.00\textwidth,keepaspectratio]{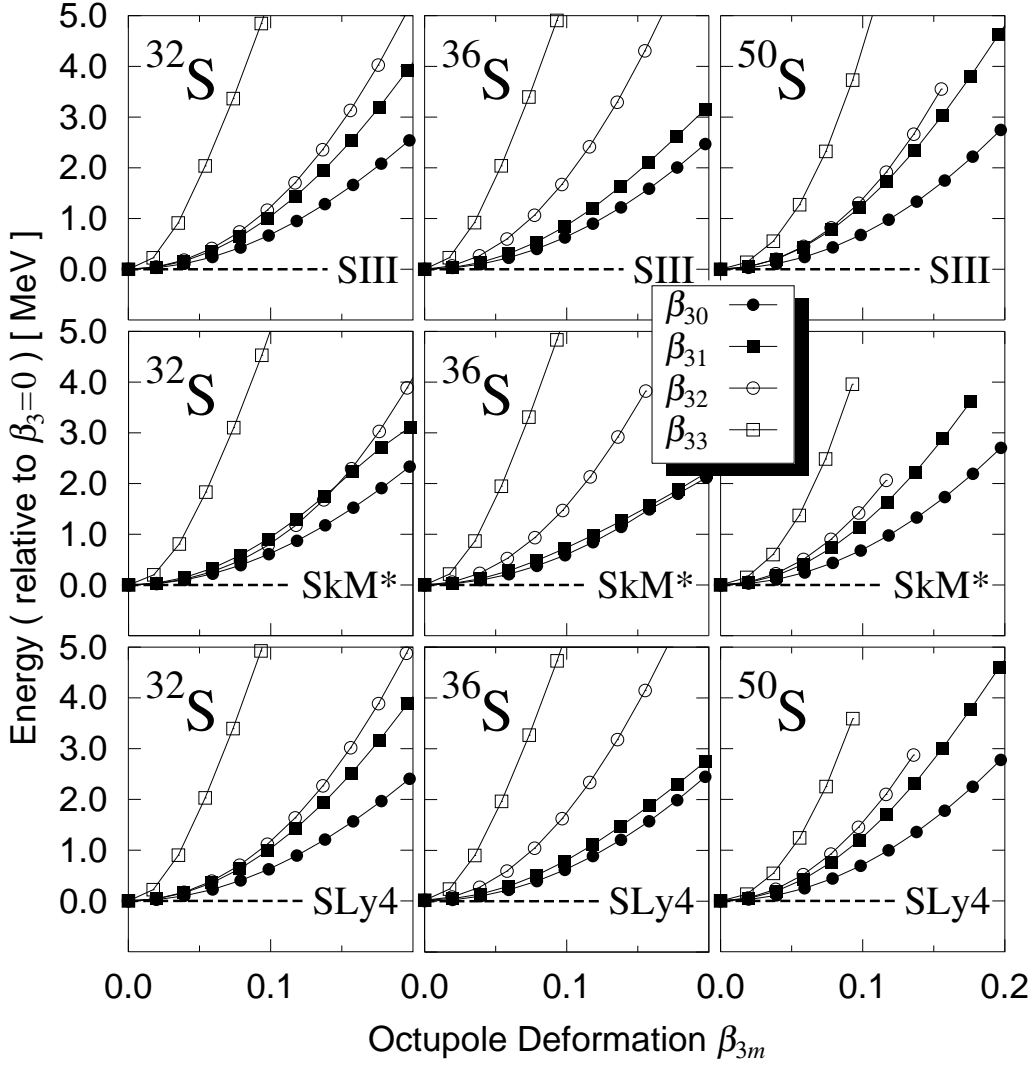}
\caption{\small
Deformation energy curves 
(measured from energies at $\beta_3=0$) as functions of 
the octupole deformation parameters  $\beta_{3m}(m=0,1,2,3)$,  
calculated for $^{32}$S, $^{36}$S, and $^{50}$S,
by means of the constrained HF procedure
with the use of the SIII, SkM$^*$ and SLy4 interactions.
The quadrupole deformation parameters are fixed at 
the equilibrium value of $\beta_2$ in each nucleus and $\gamma=0$.
One of the $\beta_{3m}(m=0,1,2,3)$ is varied while the other
$\beta_{3m}$'s are fixed to zero.
The deformation parameters $\beta_3$ and $\beta_{3m}$ are defined 
in terms of the expectation values of the octupole operators 
(see Ref.\cite{ina02a} for their explicit expressions). 
}
\end{figure}
\end{center}

\begin{center}
\begin{figure}
\includegraphics[width=1.00\textwidth,keepaspectratio]{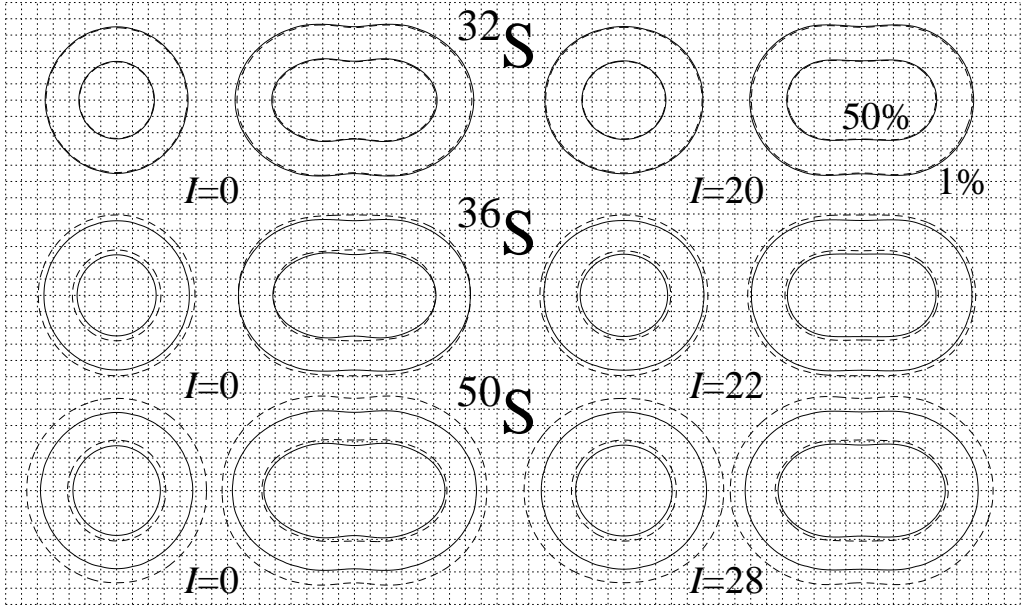}
\caption{\small
Left-hand side:
Density distributions in the $(y,x)$- and $(z,x)$- planes 
of the SD band at $I=0$
in $^{32}$S, $^{36}$S, and $^{50}$S, 
calculated with the use of the SLy4 interaction.
Neutron (proton) equi-density lines with 50\% and 1\% of
the central density are shown by dashed (solid) lines
(the inner and outer lines correspond to the 50\% and 1\% lines, 
respectively).
Right-hand side:
Same as the left-hand side but for $I=20, 22, 28$
for $^{32}$S, $^{36}$S, and $^{50}$S, respectively.
}
\end{figure}
\end{center}

\end{document}